\documentclass[aps,showpacs,preprintnumbers,amsmath,amssymb,nofootinbib]{revtex4}

\usepackage{graphicx}
\usepackage{color}

\def \be  {\begin{equation}}
\def \ee  {\end{equation}}
\def \ee  {\end{equation}}
\def \bea {\begin{eqnarray}}
\def \eea {\end{eqnarray}}

\def\be {\begin{equation}}
\def\ee {\end{equation}}
\def\bea {\begin{eqnarray}}
\def\eea {\end{eqnarray}}
\def\bc {\begin{center}}
\def\ec {\end{center}}
\def\bfg {\begin{figure}}
\def\efg {\end{figure}}
\def\bi {\begin{itemize}}
\def\ei {\end{itemize}}

\def\la {\label}
\def\le {\left}
\def\ri {\right}

\def\a  {\alpha}

\def\D  {\Delta}

\def\beq{\begin{equation}}
\def\eeq{\end{equation}}
\def\br{\begin{eqnarray}}
\def\er{\end{eqnarray}}
\newcommand{\eel}[1] {\label{#1}\end{equation}}

\begin{document}
\preprint{ECTP-2012-06}
\title{Quantum Gravity effect on the Quark-Gluon Plasma}

\author{I. Elmashad}
\email{ibrahim.elmashad@fsc.bu.edu.eg}
\affiliation{Physics Department, Faculty of Science, Benha University, Benha 13518, Egypt}

\author{A. Farag Ali}
\email{ahmed.ali@fsc.bu.edu.eg}
\email{ahmed.ali@uleth.ca}
\affiliation{Physics Department, Faculty of Science, Benha University, Benha 13518, Egypt}

\author{L. I. Abou-Salem}
\email{loutfy.Abousalem@fsc.bu.edu.eg}
\affiliation{Physics Department, Faculty of Science, Benha University, Benha 13518, Egypt}

\author{Jameel-Un Nabi} 
\email{jameel@giki.edu.pk}
\affiliation{Faculty of Engineering Sciences, GIK Institute of Engineering Sciences and Technology, Topi 23640, Khyber Pakhtunkhwa, Pakistan}
\affiliation{Egyptian Center for Theoretical Physics (ECTP), MTI University, Al-Mokattam, Cairo, Egypt}

\author{A.~Tawfik}
\email{a.tawfik@eng.mti.edu.eg}
\email{atawfik@cern.ch}
\affiliation{Egyptian Center for Theoretical Physics (ECTP), MTI University, Cairo, Egypt}
\affiliation{Research Center for Einstein Physics, Freie-University Berlin, Berlin, Germany}

\begin{abstract}
The Generalized Uncertainty Principle (GUP), which has been predicted by various theories
of quantum gravity near the Planck scale is implemented on deriving the thermodynamics of ideal Quark-Gluon Plasma (QGP) consisting of two massless quark flavors at the hadron-QGP phase equilibrium and at a vanishing chemical potential. The effective degrees of freedom and MIT bag pressure are utilized to distinguish between the hadronic and partonic phases. We find that GUP makes a non-negligible contribution to all thermodynamic quantities, especially at high temperatures.  The asymptotic behavior of corresponding QGP thermodynamic quantities characterized by the Stephan-Boltzmann limit would be approached, when the GUP approach is taken into consideration.
\end{abstract}



\maketitle


\section{Introduction}
\par\noindent
Various theories of quantum gravity predict essential modifications in the Heisenberg's uncertainty principle
near the Planck scale. In this paper, we utilize the proposed  generalized uncertainty principle (GUP), which proved 
compatible with string theory, doubly special relativity and black hole physics. It seems that this approach accordingly modifies almost all mechanical Hamiltonians. Therefore,  it can be implemented on studying the thermodynamics. 

The existence of a minimal length is one of the most interesting predictions of some approaches related to quantum gravity such as string theory as well as black hole physics. This is a consequence of string theory since strings can not interact at distances smaller than their size which yields GUP \cite{guppapers}. Black hole (BH) physics suggests that the uncertainty relation should be modified near the Planck energy scale because of measuring the photons emitted from the black hole suffers from two major errors. The first one is the error by Heisenberg classical analysis and the second one is because the black hole mass varies during the emission process and the radius of the horizon changes accordingly \cite{guppapers,BHGUP,kmm,kempf,brau}.

Recently, a new model of GUP was proposed \cite{advplb,Ali:2010yn,Das:2010zf}. It predicts a maximum observable momentum and a minimal measurable
length. Accordingly, $[x_i,x_j]=[p_i,p_j]=0$ (via the Jacobi identity) results in.
\bea
[x_i, p_j]\hspace{-1ex} &=&\hspace{-1ex} i \hbar\hspace{-0.5ex} \left[  \delta_{ij}\hspace{-0.5ex}
- \hspace{-0.5ex} \alpha \hspace{-0.5ex}  \le( p \delta_{ij} +
\frac{p_i p_j}{p} \ri)
+ \alpha^2 \hspace{-0.5ex}
\le( p^2 \delta_{ij}  + 3 p_{i} p_{j} \ri) \hspace{-0.5ex} \ri], \label{eq:alfaa}
\label{comm01}
\eea
where $\alpha = {\alpha_0}/{M_{p}c} = {\alpha_0 \ell_{p}}/{\hbar}$ and $M_{p} c^2$ stand for Planck energy. $M_{p}$ and $\ell_{p}$ is Planck mass and length, respectively. $\alpha_{0}$ sets on the upper and lower bounds to $\alpha$. Apparently, Eqs. (\ref{comm01}) imply the existence of a minimum measurable length and a maximum
measurable momentum
\bea
\D x_{min} & \approx & \alpha_0\ell_{p} \la{dxmin} , \\
\D p_{max} &\approx & \frac{M_{p}c}{\alpha_0} . \la{dpmax},
\eea
\par\noindent
where $\D x \geq \D x_{min}$ and $\D p \leq \D p_{max}$.
Accordingly, for a particle having a distant origin and an energy scale comparable to the Planck's one, the momentum would be a subject of a  modification \cite{advplb,Ali:2010yn,Das:2010zf}
\bea 
p_i = p_{0i} \le( 1 - \a p_0 + 2\a^2 p_0^2 \ri)~, \la{mom1}
\eea
where $x_i = x_{0i}$ and $p_{0j}$ satisfy the canonical commutation relations $ [x_{0i}, p_{0j}] = i \hbar~\delta_{ij}$ and simultaneously fulfil Eq. (\ref{comm01}). Here, $p_{0i}$ can be interpreted as the momentum at low energies (having the standard representation in position space, i.e. $p_{0i} = -i
\hbar \partial/\partial{x_{0i}}$) and $p_{i}$ as that at high energies.

The proposed GUP is assuming that the space is discrete, and that all measurable lengths are quantized in units of a fundamental minimum and measurable length. The latter can be as short as the Planck length \cite{advplb,Ali:2010yn}.
In order to support the idea of this procedure, we can mention that similar quantization of the length (spatial dimensions) has been studied in context of loop quantum gravity \cite{LQG}. Furthermore, it has been suggested recently \cite{Nature} that the GUP implications can be measured
directly in quantum optics lab which seems to confirm the theoretical predictions  \cite{Ali:2011fa,dvprl}

Since the GUP apparently modifies the fundamental commutator bracket between position and momentum operators, then it is natural to expect that this would result in considerable modifications in the Hamiltonian. Furthermore, it would affect a host of quantum phenomena, as well. It is important to make a quantitative study of these effects. In a series of earlier papers, the effects of GUP  was investigated on atomic and condensed matter systems \cite{dvprl,Ali:2010yn,dvcjp,Ali:2011fa}, on the weak equivalence principle (WEP), and on the Liouville theorem (LT) in statistical mechanics \cite{faragali}. For instance, it has been found that  GUP can potentially
explain the small observed violations of the WEP in neutron interferometry experiments \cite{exp}. Also, it can predict the existence of a modified invariant phase space which is relevant to the Liouville theorem.

In this paper, we present a study for the impact of the GUP on QGP consisting of two massless quark flavors at hadron-QGP phase equilibrium and at a vanishing chemical potential. We calculate the corrections to various thermodynamic quantities, like pressure. The effective degrees of freedom and MIT bag pressure are utilized in order to distinguish between the hadronic and partonic phases. 

This paper is organized as follows. In section \ref{sec:qgpthermo}, the thermodynamics of QGP consisting of two massless quark flavors is briefly reviewed. In section \ref{sec:GUPQGP}, the effects GUP \cite{advplb, Das:2010zf} on QGP thermodynamics is outlined. The results and discussions are given section \ref{sec:rslt}.

\section{Thermodynamics of Quark-Gluon Plasma}
\label{sec:qgpthermo}

In this section, we briefly review the thermodynamics of QGP \cite{QGP}.
At finite temperature $T$ and chemical potential $\mu$, the grand-canonical partition function $z_{B}$ for non-interacting massive bosons with $g$ internal degrees of freedom is given as
\bea
\ln z_{B}=\prod_{k} \le[\sum_{l=0}^{\infty} \exp\le(-l\frac{E(k)-\mu}{T}\ri)\ri]^{g}
         =\prod_{k} \le[1-\exp{(E(k)-\mu)}\ri]^{-g},
\eea
where $k$ is the momentum of the particle and $l$ is the occupation number for each quantum state with energy $E(k)=\sqrt{k^{2}+m^{2}}$ with mass $m$. Here the infinite product is taken for all possible momentum states. For simplicity we consider a massless pion gas in a vanishing chemical potential.  Then the grand potential reads
\bea \label{eq:pf1}
\dfrac{\Omega}{V}=g \int_{0}^{\infty}\dfrac{ d^{3}k}{(2\pi)^{3}}T\ln\le[1-\exp{\dfrac{-E(k)}{T}}\ri]  = -g \dfrac{\pi^{2}}{90}T^{4}
\eea
Therefore, pressure $P$, energy density  $\varepsilon$, and entropy density $S$ of hadronic state can be deduced
\begin{eqnarray}
P_{H} &=& g_{\pi}\dfrac{\pi^{2}}{90}T^{4}, \\
\varepsilon_{H} &=& 3 g_{\pi}\dfrac{\pi^{2}}{90}T^{4}, \\
S_{H} &=& 4 g_{\pi}\dfrac{\pi^{2}}{90}T^{3}.
\end{eqnarray}

Using MIT bag model \cite{mitbag}, then the QGP thermodynamic quantities accordingly read
\begin{eqnarray}
P_{QGP} &=& g_{QGP}\dfrac{\pi^{2}}{90}T^{4}-B, \label{eq:pqgp1}\\
\varepsilon_{QGP} &=& 3 g_{QGP}\dfrac{\pi^{2}}{90}T^{4}+B, \label{eq:eqgp1} \\
S_{QGP} &=& 4 g_{QGP}\dfrac{\pi^{2}}{90}T^{3}, \label{eq:sqgp1}
\end{eqnarray}
where $B$ is the bag pressure. Based on Gibbs condition \cite{gibbs}, the critical point is obtained at the phase equilibrium, which is satisfied when $P_{H}(T_{c})= P_{QGP}(T_{c})$. Then, the bag pressure can given as
\be
 B=(g_{QGP}-g_{\pi})\dfrac{\pi^{2}}{90}T_{c}^{4}
\ee

\section{GUP and QGP}
\label{sec:GUPQGP}

For a particle of mass $ M $ having a distant origin and an energy  comparable to the Planck scale, the momentum would be a subject of a tiny modification and so that the dispersion relation would too \cite{advplb,Das:2010zf,Ali:2011fa,Tawfik:2012hz}
\begin{equation}
E(k)^{2}=k^{2}c^{2}(1-2\alpha {k})+M^{2}c^{4},
\end{equation}
where $c$ and $M_{pl}$ are the speed of light as introduced by Lorentz and implemented in special relativity and Planck's mass, respectively. For simplicity we use  natural units in which $\hbar=c=1$ and will consider a massless pion gas
\begin{equation}
E(k)=k(1-2\alpha {k})^{1/2}
\end{equation}
For large volume, the sum over all states of single particle can be rewritten in terms of an integral \cite{faragali}
\begin{equation}
\sum_{k}\rightarrow{\dfrac{V}{(2\pi)^{3}}\int_{0}^{\infty}d^{3}k}\rightarrow{\dfrac{V}{2\pi^{2}}\int_{0}^{\infty}\dfrac{k^{2}dk}{(1-\alpha{k})^{4}}}
\end{equation}
Therefore, the partition function, Eq. (\ref{eq:pf1}), reads
\begin{eqnarray}
\ln z_{B}&=& -\dfrac{V g}{2 \pi^{2}} \int_{0}^{\infty} k^{2} \frac{\ln \le[1-\exp\le(-\frac{E(k)}{T}\ri)\ri]}{(1-\alpha k)^{4}} ~dk
         = -\dfrac{V g}{2 \pi^{2}} \int_{0}^{\infty}  k^{2} \frac{\ln \le[1-\exp\le(-\dfrac{k}{T}(1-2 \alpha k)^{1/2}\ri)\ri]}{(1-\alpha k)^{4}} ~dk, \nonumber \\
 &=&\dfrac{-V g}{2\pi^{2}}\le[\le.\dfrac{k^{3}}{3(1-\alpha k)^{3}}\ln \le[1-\exp\le(-\dfrac{k}{T}(1-2 \alpha k)^{1/2}\ri) \ri]\ri|_{0}^{\infty} \right. \label{eq:lnaaa} \hspace*{5mm} \\
 &-& \left.  
 \int_{0}^{\infty}\dfrac{k^{3}}{3(1-\alpha k)^{3}}\dfrac{\dfrac{1}{T} \exp\le(-\dfrac{k}{T}(1-2 \alpha k)^{1/2}\ri)}{1-\exp\le(-\dfrac{k}{T}(1-2 \alpha k)^{1/2}\ri)}\le[\dfrac{1-3\alpha k}{(1-2\alpha k)^{1/2}}\ri] dk\ri]. \label{eq:lnbbb}
\end{eqnarray}
It is obvious that the first term in Eq. (\ref{eq:lnbbb}) vanishes. Thus
\be
\ln z_{B}= \dfrac{V g}{2\pi^{2}}\int_{0}^{\infty}\dfrac{k^{3}}{3(1-\alpha k)^{3}}\dfrac{\dfrac{1}{T}\le[\dfrac{1-3\alpha k}{(1-2\alpha k)^{1/2}}\ri] }{\exp\le(-\dfrac{k}{T}(1-2 \alpha k)^{1/2}\ri)-1} dk.
\ee

Let $x=\dfrac{k}{T}(1-2 \alpha k)^{1/2}$~ so that $dx=\dfrac{1}{T}\dfrac{1-3\alpha k}{(1-2\alpha k)^{1/2}} dk$ and 
\begin{equation}
\ln z_{B}=\dfrac{V g}{2\pi^{2}}\int_{0}^{\infty}\dfrac{k^{3}}{3(1-\alpha k)^{3}}\dfrac{dx}{e^{(x)}-1}
\end{equation}
Apparently, as we are interested in the terms containing the first order of $\alpha$, so $x$ can be approximated as follows.
\bea
x&=&\dfrac{k}{T}(1-2\alpha k)^{1/2} \approx \dfrac{k(1-\alpha k)}{T},
\eea
and therefore
\bea
k&=& x T + \alpha k^2 \approx x T + \alpha  (x^2 T^2 + 2 \alpha x T k^2 + \alpha^2 k^4) \la{xT}.
\eea
When ignoring higher orders  of $\alpha$, then
\be
k\approx xT(1+\alpha{xT}).
\ee

The partition function becomes
\begin{equation}
\ln z_{B}=\dfrac{V g}{2\pi^{2}}\int_{0}^{\infty} \frac{1}{3} x^{3}T^{3} \dfrac{(1+\alpha xT)^{3}}{(1-\alpha xT)^{3}}\dfrac{dx}{e^{x}-1}.
\end{equation}
It is apparent that the integral contains Maclaurin series, which is a Taylor series expansion of a function about $0$  
\begin{equation}
\dfrac {(1+\alpha xT)^{3}}{(1-\alpha xT)^{3}}=1+6\alpha{xT}+[\cdots] \alpha^{2}x^{2}T^{2}+[\cdots] \alpha^{3}x^{3}T^{3}+.... ,
\end{equation}
where $[\cdots]$ stands for non-identical factors. 
When ignoring all terms cantoning $\alpha$ with order $\geq 2$, then 
\begin{equation}
\ln z_{B}=\dfrac{V g}{6\pi^{2}}\le[\int_{0}^{\infty} T^{3} \dfrac{ x^{3} dx}{e^{x}-1}+ \int_{0}^{\infty} 6 \alpha T^{4} \dfrac{ x^{4} dx}{e^{x}-1}\ri].
\end{equation}
The partition function is related to the grand canonical potential, $\ln z_{B}=-\Omega/T$,
\begin{eqnarray}
\dfrac{\Omega}{V} &=& -\dfrac{g}{6\pi^{2}} T^{4} [\Gamma{(5)}I(0)_{5}^{-}] - \alpha \dfrac{g}{\pi^{2}} T^{5} [\Gamma{(6)}I(0)_{6}^{-}] \la{omega},
\end{eqnarray}
where$I(0)_{n}^{\pm}$ is  Bose and Fermi integrals
\begin{equation}
I(y)_{n+1}^{\pm}=\dfrac{1}{\Gamma{(n+1)}}\int_{0}^{\infty} \dfrac {x^n}{(x^{2}+y^{2})^{1/2}}\dfrac{1}{\exp{\left[(x^{2}+y^{2})^{1/2}\right]}\pm{1}} \la{IY} dx,
\end{equation}
and $\Gamma{(n)}=(n-1)!$ is the gamma function. At $ y=0 $, then 
\begin{equation}
I(0)_{n+1}^{\pm}=\dfrac{1}{n}\zeta{(n)}a_{n}^{\pm} \la{I0}
\end{equation}
This formula is valid for integer ${(n\geq{2})}$ with $a_{n}^{+}= 1-2^{(1-n)},a_{n}^{-}=1$. The Riemann zeta function, $ \zeta{(n)}=\sum_{j=1}^{\infty} j^{(-n)} $ with $\zeta{(2)}=\dfrac{\pi^2}{6},\zeta{(3)}=1.202\zeta{(4)}=\dfrac{\pi^2}{90},\zeta{(5)}=1.037$, etc.
Substituting from Eq. (\ref{IY}) and Eq. (\ref{I0}) into Eq. (\ref{omega}) we have,
\be
\dfrac{\Omega}{V}=-g\dfrac{\pi^{2}}{90} T^{4} - 24\alpha\dfrac{g}{\pi^{2}}  T^{5}\zeta{(5)}
                 =-g\dfrac{\pi^{2}}{90}T^{4}- g\alpha_{1}T^{5},
\ee
where $ \alpha_{1}=24\alpha\zeta{(5)}/\pi^{2}$.

The pressure is directly related to the grand canonical potential, $P=-\Omega/V$. Then, in the hadronic phase
\be \label{eq:ppG}
P_H = g\dfrac{\pi^{2}}{90}T^{4} + g\alpha_{1}T^{5}
\ee
In a massless ideal hadronic gas, the energy density is related with the pressure by the equation of state, $\varepsilon_H=3\, P_H$,  and the entropy density is given by the derivative of pressure with respect to temperature $T$
\begin{eqnarray}
\varepsilon_H &=& 3 g\dfrac{\pi^{2}}{90}T^{4} + 3 g \alpha_{1}T^{5}, \label{eq:eeG}\\
S_H &=& 4 g \dfrac{\pi^{2}}{90}T^{3} + 5 g \alpha_{1}T^{4}. \label{eq:ssG}
\end{eqnarray}
Taking into consideration the relevant degrees of freedom, it is obvious that this set of equations, Eq. (\ref{eq:ppG}),  (\ref{eq:eeG}) and (\ref{eq:ssG}), is valid in the hadronic state. In QGP state, the bag pressure should be inserted as done in Eq. (\ref{eq:pqgp1}) and (\ref{eq:eqgp1}).
From Gibbs phase equilibrium condition $P_{H}(T_{c})= P_{QGP}(T_{c})$ \cite{gibbs}, the bag pressure $B$ can be deduced
\begin{equation}
\dfrac{\pi^{2}}{90}T_{c}^{4}+\alpha_{1}T_{c}^{5}=\dfrac{B}{g_{QGP}-g_{\pi}} \la{TC}.
\end{equation}
Then, at this value for the bag pressure, the thermodynamic pressure in QGP state reads
\begin{equation} \label{eq:Pp}
\dfrac{P_{QGP}}{T^{4}}=g_{QGP}\dfrac{\pi^{2}}{90}\le[1-\dfrac{g_{QGP}-g_{\pi}}{g_{QGP}}\le(\dfrac{T_{c}}{T}\ri)^{4}\ri]+g_{QGP}\alpha_{1}T\le[1-\dfrac{g_{QGP}-g_{\pi}}{g_{QGP}}\le(\dfrac{T_{c}}{T}\ri)^5\ri],
\end{equation}
from which the normalized QGP energy density can be written as 
\begin{eqnarray}  \label{eq:Ee}
\dfrac{\varepsilon_{QGP}}{T^{4}} &=&  g_{QGP}\dfrac{\pi^{2}}{90}\le[3+\dfrac{g_{QGP}-g_{\pi}}{g_{QGP}}\le(\dfrac{T_{c}}{T}\ri)^{4}\ri] + g_{QGP}\alpha_{1}T\le[1+\dfrac{g_{QGP}-g_{\pi}}{g_{QGP}}\le(\dfrac{T_{c}}{T}\ri)^5\ri]. \end{eqnarray}
As given in Eq. (\ref{eq:sqgp1}), the QGP entropy is not depending on the bag pressure,
\begin{eqnarray}
\frac{S_{QGP}}{T^{3}} &=& 4 g_{QGP}\dfrac{\pi^{2}}{90}+5g_{QGP}\alpha_{1}T \label{eq:Ss}
\end{eqnarray}

\section{Results and discussions}
\label{sec:rslt}

\begin{figure}[htb!]
\includegraphics[width=8cm,angle=-90]{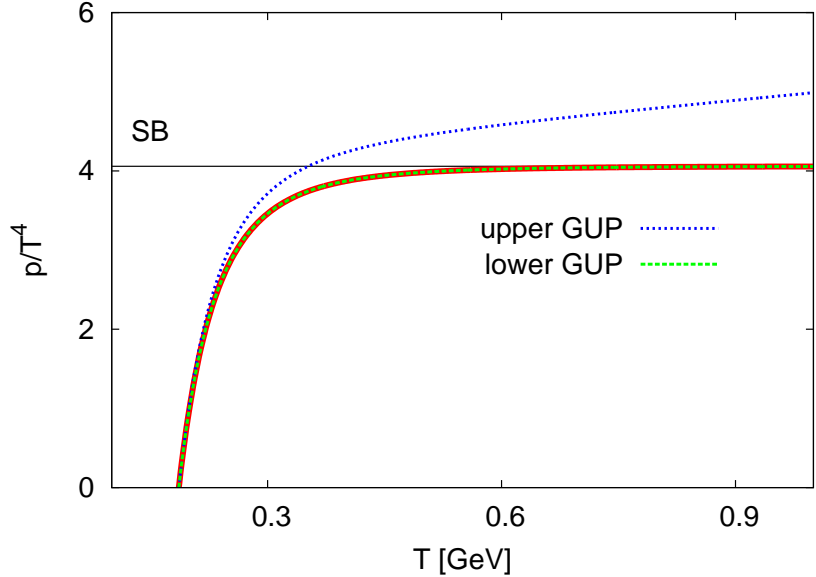}
\caption{The pressure normalized to $T^4$ is given in dependence on $T$. The horizontal solid line represents the Stephan-Boltzmann (SB) limit. The solid curve gives the result of the first term, Eq. (\ref{eq:Pp}), in which the GUP effects is not included.  The second term, the term including $\alpha_1$, is drawn by dashed and dotted curves. These two curves are distinguished by the upper and lower bounds of $\alpha$.}
\label{afig1a}
\end{figure}

In Fig. \ref{afig1a} the QGP pressure is given as function of $T$. The degrees of freedom in the QGP state assuming $n_{f}=2$ and $n_{c}=3$ are $g_{QGP}=37$. In the pionic (hadronic) state, they are $g_{\pi}=3$. The horizontal solid lines represent the Stephan-Boltzmann (SB) limits of the pressure. The solid curves give the results of the first terms, Eq. (\ref{eq:Pp}), in which the GUP effects is not included. The second term, the term including $\alpha_1$, is drawn by dashed and dotted curves. These two curves are distinguished by the values of upper and lower bounds of $\alpha$. As given in \cite{Ali:2011fa} and Eq. (\ref{eq:alfaa}), the first bound for $\alpha_0$ is about  $\sim10^{ 17}$, which would approximately gives $\alpha\sim 10^{-2}~$GeV$^{-1}$. The other bound  of $\alpha_0$  which is $\sim10^{10}$. This lower bound means that $\alpha\sim10^{-9}~$GeV$^{-1}$. As discussed in \cite{Tawfik:2012hz}, the exact bound on $\alpha$ can be obtained by comparing with observations and experiments \cite{ref17}. It seems that the gamma rays burst would allow us to set an upper value for the GUP-charactering parameter $\alpha$ \cite{tma}.

\begin{figure}[htb!]
\includegraphics[width=7cm]{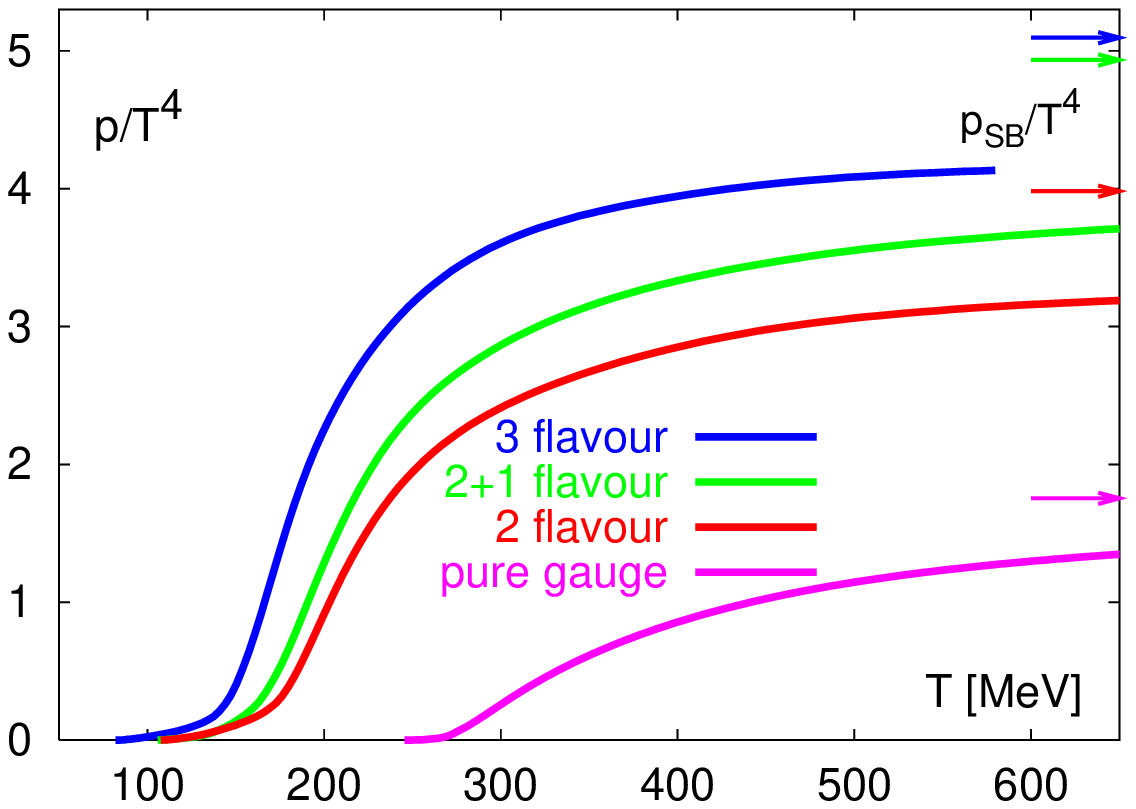}
\includegraphics[width=8.1cm,bb=18 360 592 718]{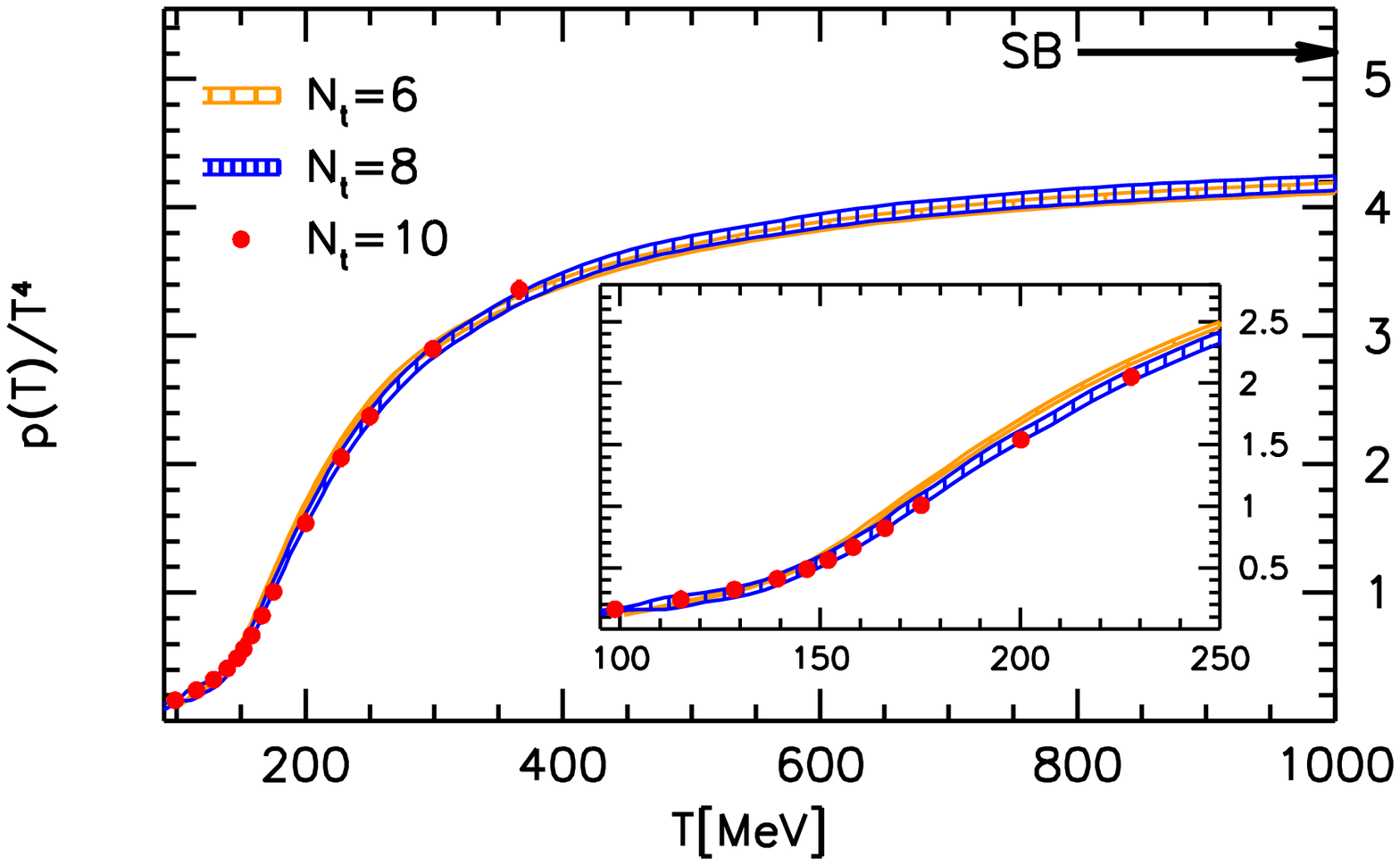}
\caption{Left panel shows the results the normalized pressure with the p4 action for different quark flavors \cite{lttcc}. The data are obtained in a very coarse lattice (small temporal dimension $N_{\tau}=4$) with bare masses $m_q/T=0.4$ and $m_s/T=1$. Right panel gives the results for $N_{\tau}=6,8$ and $10$ lattices for three quark flavors \cite{fodor2010a}. Therefore, the SB limit reads $p_{SB}/T^4 \approx 5.21$. It is indicated by an arrow. 
}
\label{afig2a}
\end{figure}

In right panel of Fig. \ref{afig2a}, the lattice QCD results for $p/T^4$ for dynamic three ($n_f=3$) quarks are given in dependence on $T$ and for different temporal dimensions of the lattice  \cite{fodor2010a}.  The lattice temporal dimension $N_{\tau}$ is directly related to the inverse temperature, $1/T=N_{\tau}\, a$, where $a$ is the lattice spacing. We notice that for lattice QCD highest temperature $T=1000$ MeV, the pressure, for instance, is almost $20\%$ below the SB limit. Other lattice QCD simulations for different quark flavors seem to confirm that the thermodynamics of massive quarks is below SB \cite{karsch2009a,lttcc}, left panel of Fig. \ref{afig2a}. Various studies for the lattice QCD thermodynamics with different quark masses and flavors have been performed over the last $2-3$ decades. For a recent review on the lattice QCD equation of state, the readers are referred to \cite{millerA,owe}. In all these studies, the resulting thermodynamic quantities above $T_c$ are found to be smaller than the SB limit of the quantity of interest. For a long time, this observation stands without a clear interpretation or at least the worldwide community was not unified on one single interpretation. At that time, the idea that hadrons above $T_c$ is conjectured to be dissolved into non-correlated almost independent constituents of quarks and gluons was widely distributed. The latter are believed to for an ideal QGP. With the RHIC discovery \cite{rhic2005}, an equilibrated, but strongly coupled QGP is formed in the heavy-ion collisions. Therefore, the distinguish of its thermodynamics from SB turns to be eligible. It is widely believed that the SB limit can be approached when the strong running coupling $\alpha_s$ nearly vanishes. For a recent experimental evaluation of $\alpha_s$, the readers are referred to \cite{alfas}. In light of lattice QCD thermodynamics and $\alpha_s(Q^2)$ and in case of neglecting the effects of quantum gravity, the partonic constituents would form an ideal gas at very high temperatures or energies. If the GUP approach is taken into consideration, the asymptotic behavior that is characterized by the SB limit would be reached than the earlier case. Based on this study, one would feel encouraged to implement GUP approach on QGP with dynamic thee quarks and having strong correlations and dissipative properties \cite{Tawfik:2011gh,Tawfik:2011sh,Tawfik:2011mw,Tawfik:2010pm,Tawfik:2010mb,Tawfik:2010bm,Tawfik:2010ht,Tawfik:2009nh,Tawfik:2009mk}.

\section*{Acknowledgments}
The research of IE,  AFA,  and LIA-S is supported by Benha University. The research of AT has been partly supported by the German--Egyptian Scientific Projects (GESP ID: 1378). AT likes to thank Prof. Antonino Zichichi for his kind invitation to attend the twenty-ninth World Laboratory Meeting at the ''Ettore Majorana Foundation and Centre for Scientific Culture'' in Erice-Italy, where the present work is completed. 
 

\appendix
\section{Integration by parts for Eq. (\ref{eq:lnaaa})}

In order to solve Eq. (\ref{eq:lnaaa}), let us assume that
\begin{eqnarray}
u&=&\ln \le(1-e^{-\dfrac{k}{T}(1-2 \alpha k)^{1/2}} \ri) \hspace{10mm} \mathtt{and} \hspace{10mm}
dv = \dfrac{k^{2}}{(1-\alpha k)^{4}}.
\end{eqnarray}
Then
\begin{eqnarray}
d u &=&\dfrac{\dfrac{1}{T}e^{-\dfrac{k}{T}(1-2 \alpha k)^{1/2}}}{1-e^{-\dfrac{k}{T}(1-2 \alpha k)^{1/2}}}\le[\dfrac{1-3\alpha k}{(1-2\alpha k)^{1/2}}\ri] dk, \hspace{10mm} \mathtt{and} \hspace{10mm}
v =\dfrac{k^{3}}{3(1-\alpha k)^{3}}.
\end{eqnarray}
The partition function, Eq. (\ref{eq:lnaaa}) can be re-written as
\be
\ln z_{B} = -\frac{d}{2\pi^{2}} V \, \le.uv\ri|_{0}^{\infty} + \frac{d}{2\pi^{2}}\, V  \int_{0}^{\infty}v\, du.
\ee

%
%

\end{document}